\documentclass[prl,aps,amssymb,showpacs,twocolumn]{revtex4}
\usepackage{amsmath}
\usepackage{amssymb}
\usepackage{amsthm}
\usepackage{amsfonts}
\usepackage{enumerate}
\usepackage{latexsym}
\usepackage[dvips]{graphicx}

\newcommand{\beq}{\begin{equation}}
\newcommand{\eneq}{\end{equation}}

\def\be{\begin{equation}}       \def\ee{\end{equation}}
\def\bea{\begin{eqnarray}}      \def\eea{\end{eqnarray}}

\input{epsf}

\begin{document}

\tolerance 10000

\newcommand{\vk}{{\bf k}}

%\draft

\title{ Topological Orbital Angular Momentum Hall Current}
\author {  Jiangping Hu   }

\affiliation{  Department of  Physics, Purdue University,
West Lafayette, IN 47907\\
        }
\begin{abstract}
We show that there is a fundamental difference between spin Hall
current and orbital angular momentum Hall current in Rashba-
Dresselhaus spin orbit coupling systems. The orbital angular
momentum Hall current has a pure topological contribution which is
originated from the existence of magnetic flux in momentum space
while there is no such topological nature for the spin Hall
current. Moreover, we show that the  orbital Hall conductance is
always larger than the spin Hall conductance in the presence of
both couplings. The topological part is expected to be free from
the effect of disorder due to the topological nature. Therefore,
the orbital angular momentum Hall current should be the major
effect in real experiments.
\end{abstract}


\maketitle

The intrinsic spin Hall effect\cite{Murakami2003,Sinova2004} in spin
orbit coupling systems has attracted intensive research attentions
recently. The effect has been studied in a broad class of spin-orbit
coupling models, such as Luttinger, Rashba and Dresselhaus
Hamiltonians \cite{Hu2003,Bernevig2004a, Bernevig2005a,
Culcer2004a,Rashba2003,Murakami2004a,
Murakami2004b,Murakami2004c,Rashba2004,Bernevig2004b,Burkov2004a,Chang2004,Hirsch2004,
Schliemann2003,Schliemann2004a,Sinitsyn2004a,Shen2004a,Sheng2004b,Liu2004}.
Although the spin Hall effect seems to be universal in strong spin
orbit coupling systems and even most recently, two experimental
groups \cite{Kato2004,Wunderlich2004} have reported that their
experimental results are consistent with the physics of the spin
Hall effect predicted in two dimensional electron systems, it is
still a fundamental question regarding whether the effect is a true
``observable" effect.

There are at least two issues regarding this observability. The
first is the role of the disorder. Several theoretical analyses
have shown that the spin Hall effect in the Rashba or Dresselhaus
systems disappears in the presence of
disorder\cite{Mishchenko2004,Inoue2004,Liu2005}, even in the weak
disorder limit. These results suggest that the spin Hall effect is
only measurable in the ballistic regime. The second question is
even more fundamental. In the spin orbit coupling systems, it is
improper to define a pure spin current since the orbit and spin
are coupled. It is the total angular momentum which is the true
observable quantity. This point has been raised by Zhang and
Yang\cite{Zhang2004a}. They have argued that the intrinsic spin
Hall current is always accompanied by an equal but opposite
intrinsic orbital angular momentum Hall current. Therefore, the
intrinsic spin Hall effect does not induce  spin accumulation at
the edge of the sample. This argument leads to a conclusion that
there is no measurable effect induced by the spin Hall current.

Besides the two issues, there are also issues which are related to
the proper definition of the spin current. Since spins are not
conserved in spin orbit coupling systems, a conserved quantity may
be needed to replace the spin operator in order to define a
conserved spin current. In the Luttinger spin orbit coupling
systems, such formulations have been proposed\cite{Murakami2004a}.
The conserved spin current has a very beautiful physical
interpretation. The origin of the current is from the existence of
a monopole structure in momentum space, which are directly derived
from the Berry phase in spin orbit coupling
systems\cite{Murakami2003,Hu2003} when Hamiltonian is projected to
double degenerated helicity bands. Due to this topological nature,
the conserved spin Hall current in general can be called as
topological spin Hall current. The same topological nature also
accounts for the anomalous Hall
effect\cite{Karplus1954,Jungwirth2002} in ferromagnetic metal,
which has been confirmed in experimental\cite{Lee2004} and
numerical studies\cite{Fang2003}. However, the topological spin
Hall current is not universal for other spin orbit coupling
systems. In the Rashba or Dresselhaus spin orbit coupling systems,
the topological spin Hall current is zero\cite{Hu2003}. The reason
is that in the Luttinger case, the spin operator after being
projected to each band is nontrivial while it disappears in the
Rashba and Dresselhaus case.

The topological part of the current can be viewed as an intrinsic
property associated with an individual band. It is protected by
the topological nature and is insensitive to the effect of the
weak disorder. In this letter, we study the orbital angular
momentum Hall current in the presence of both Rashba and
Dresselhaus spin orbit coupling. In this model, the orbital Hall
current has a very different origin from the spin Hall current.
The orbital Hall current has a fundamental topological nature. It
is originated from the existence of the magnetic flux in momentum
space while there is no such topological nature for the spin Hall
current. Contradictory to the result in ref.\cite{Zhang2004a}, the
orbital Hall current does not cancel the spin Hall current in
general and the orbital Hall conductance does not change sign for
any coupling parameters while it has been shown that when the
Dresselhaus coupling is larger than the Rashba coupling strength,
the spin Hall conductance changes the
sign\cite{Sinitsyn2004a,Shen2004a}. In fact, the orbital Hall
current is larger than the spin current in the presence of both
couplings. Therefore, the total angular momentum current in
general is dominated by the orbital Hall current and is
non-vanishing. The intrinsic total angular momentum Hall effect
does generate magnetization at the edge of samples. Moreover, the
orbital current is free from the effect of impurity due to the
topological nature. From these results, we argue that the orbital
Hall current dominates the transport properties in such systems.

We consider a Hamiltonian for a two dimensional electron system
with both Rashba and Dresselhaus spin orbit coupling
\begin{equation}
\hat{H}={\frac{p^2}{2m}}+H_R+H_D,
\end{equation}
where the Rashba term is
\begin{equation}
H_R=\alpha(\vec{\sigma}\times \vec{p})\cdot\hat{z}=\alpha(p_y
\sigma_x-p_x \sigma_y)
\end{equation}
and the Dresselhaus term is
\begin{equation}
H_D=\beta(p_y \sigma_y-p_x \sigma_x).
\end{equation}
The Hamiltonian can be diagonalized by a unitary matrix
\begin{equation}
U={\frac{1}{\sqrt{2}}}\left(\begin{array}{cr}-\imath
e^{-\imath\theta}&\imath e^{-\imath\theta}\\1&1\end{array}\right)
\end{equation}
with $\tan{\theta}={\frac{\alpha p_y-\beta p_x}{\alpha p_x-\beta
p_y}}$, and
\begin{equation}
\label{diag} U^{\dag}HU={\frac{p^2}{2m}}-E_p\sigma_z.
\end{equation}
with
\begin{equation}
E_p=\sqrt{(\alpha^2+\beta^2)p^2-4\alpha\beta p_x p_y},
\end{equation}
which is the band gap due to the spin orbit coupling.

First, we follow the standard definition of the current. Similar
to the spin current, we can define the orbital angular momentum
current as
\begin{eqnarray}
O_j^i=\frac{1}{2}\{L_i,  v_j\}
\end{eqnarray}
where $L_i=\epsilon_{ijk}x_jp_k$ are the orbital angular momentum
operators and $v_j= \frac{\partial H}{\partial p_j}$ are velocity
operators.

By a straightforward calculation following the Kubo formula in the
ballistic regime which is given by
\begin{eqnarray}
&
&\sigma_{yx}^{L_z}=e\sum_{\lambda,\lambda'\neq\lambda}\int{\frac{d\vec{p}}{(2\pi)^2}}(f_{\lambda',p}-f_{\lambda,p})
\nonumber \\
& &{\frac{\textrm{Im}[\langle
\lambda',p|O_{y}^{z}|\lambda,p\rangle\langle\lambda,p|v_{x}|\lambda',p\rangle]}{(E_{p\lambda'}-E_{p\lambda})^2}},
\end{eqnarray}
the orbital angular momentum Hall current is calculated to be
\begin{eqnarray}
O^z_y=
\sigma_{yx}^{L_z}E_x=\frac{e}{8\pi}\frac{\alpha^2+\beta^2}{|\alpha^2-\beta^2|}
E_x
\end{eqnarray}
for $\alpha\neq\beta$, where we assume the applied electric field
is along the $x$ direction. For $\alpha=\beta$, $O^z_y=0$. One can
compare this result with the spin Hall current, which is given
by\cite{Sinitsyn2004a,Shen2004a}
\begin{eqnarray}
\frac{J_y^z}{E_x}=\left \{ \begin{array}{cc} -\frac{e}{8\pi} & \alpha>\beta \\
0 & \alpha=\beta \\
\frac{e}{8\pi} & \alpha<\beta
\end{array} \right.
\end{eqnarray}
First, we notice that the orbital conductance has the same
absolute value as the spin conductance but with opposite sign when
only the Rashba coupling is present, i.e. $\beta=0$. This result
is obtained in ref.\cite{Zhang2004a}, which can be easily
understood because $L_z+S_z$ is conserved when $\beta=0$.
Secondly, the orbital conductance is the same as the spin
conductance when only the Dresselhaus coupling is present. This
result shows that the total angular momentum Hall current is
non-vanishing in the system. The result can also be understood as
follows. Since $L_z-S_z$ is conserved when $\alpha=0$, the orbital
and spin current must be equal. Thirdly, both spin and orbital
conductances are discontinuous across $\alpha=\beta$. The orbital
conductance remains non-negative for any spin orbit coupling
parameters while the spin conductance changes the sign across
$\alpha=\beta$. Fourthly, the absolute value of the spin
conductance is universal in the presence of the spin orbit
coupling while the orbital conductance is dependent on the
coupling strength. In fact, the value of the orbital conductance
is always larger than or equal to the spin conductance. Therefore,
the total angular momentum current is always dominated by its
orbital part. In a system with the larger value of the Rashba
coupling than the Dresselhaus coupling, this result suggests that
the direction of magnetization at the edge of the sample due to
the flowing  angular momentum Hall current should be opposite to
the previous predictions based on the pure spin Hall current.

Strictly speaking, the result that one obtains from the Kubo
formula is not an intrinsic property of the individual band. The
spin and orbital Hall currents are not conserved quantities. It is
also easy to see that in the Kubo formulism, both bands are
required. In ref.\cite{Murakami2003,Murakami2004a}, the authors
show that a conserved spin Hall current can be defined in the
Luttinger spin orbit coupling systems. In this definition, the
spin Hall current becomes an intrinsic property of the individual
band, which can be derived from an effective Hamiltonian
associated with the individual band. Moreover, the current has a
fundamental topological nature which entirely comes from the Berry
phase, which can be viewed as a monopole in momentum
space\cite{Murakami2003}. In the Rashba-Dresselhaus spin orbit
coupling systems, such a definition of the spin current does not
exist. In fact, the topological contribution to the spin Hall
current is zero. However, in the following, we will show that the
orbital angular momentum Hall conductance can come from a
fundamental topological nature due to the spin orbit coupling.

When the spin orbit coupling is very large, the independent
effective Hamiltonian should exist for each individual band. Such
Hamiltonian can be constructed by projecting the total Hilbert
space in the original model to the space in the individual band.
For the Rashba-Dresselhaus model, the effective Hamiltonians for
two bands after projection are respectively given by
\begin{eqnarray}
& &H_1=\frac{p^2}{2m}-E_p \\
& & H_2=\frac{p^2}{2m}+E_p
\end{eqnarray}
As shown in ref.\cite{Murakami2003}, such a projection introduces
a nontrivial gauge potential in momentum space. In our case, the
gauge potential $\vec A$ is given by
\begin{equation}
\vec A= \langle L_z\rangle \vec A_0,
\end{equation}
where $\langle L_z \rangle$ is the value of orbital angular
momentum in the single particle state, which is given by
\begin{equation}
\langle L_z\rangle =-\frac{(\alpha^2-\beta^2)p^2}{2E_p^2}.
\end{equation}
and $\vec A_0=(\frac{p_y}{p^2},-\frac{p_x}{p^2})$. Notice that the
gauge structure and the value of $\langle L_z \rangle$ are identical
for both bands. Physically, this gauge describes an angular momentum
flux with the value equal to $\langle L_z \rangle$ at the origin of
the momentum space, i.e. $p=0$. The coordinate operators which are
the derivatives with respect to momentum are modified to the
covariant derivatives, namely
\begin{equation}
x= i\frac{\partial}{\partial p_{x}} +A_x, y=
i\frac{\partial}{\partial p_{y}} +A_y
\end{equation}
In the Luttinger model, It has been shown that the gauge generates
a noncommutative geometry\cite{Murakami2003}. In this case, the
commutation relation, $[x,y]$, is zero everywhere except at the
origin. However, the integral in an area which includes the origin
of the momentum space is non-vanishing, namely,
\begin{equation}
\frac{1}{4\pi^2}\int [x,y] dp_xdp_y = i\frac{1}{4\pi^2}\oint \vec
A \cdot d\vec p \neq 0
\end{equation}

In the presence of the electric field, the total effective
Hamiltonian for the individual bands are
\begin{equation}
H^{1,2}_{eff}=H_{1,2}+ex E_x
 \end{equation}
Thus the orbital angular momentum Hall current is given by
\begin{eqnarray}
& &\frac{O^{z}_y}{E_x} = -i\frac{e}{4\pi^2}\int dp_xdp_y
\langle L_z\rangle [x,y] \nonumber \\
& & =\frac{e}{4\pi^2}\oint  \langle L_z \rangle ^2 \vec A_0\cdot
d\vec p
\end{eqnarray}
The above integral vanishes when the Fermi surface does not
contain the origin. When the Fermi energy is larger than zero, the
integral is nonzero. Since the Fermi surfaces in both bands
contain zero, Therefore the total topological orbital conductance
in the system is doubled. Therefore,
\begin{eqnarray}
 \frac{O^{z}_y}{E_x} =\left\{\begin{array}{ll}
 0  & E_F <0 \\
 \frac{e}{4\pi}\frac{\alpha^2+\beta^2}{|\alpha^2-\beta^2|}  & E_F>0
 \end{array}\right.
\end{eqnarray}
Where $E_F$ is the Fermi energy. The topological orbital
conductance is slightly different from the previous result derived
from the Kubo formula for the ordinary orbital conductance, which
is also the case for the topological spin current in the Luttinger
case\cite{Murakami2003,Hu2003}. This is very natural. In fact, one
can easily derive different contributions to orbital Hall current
based on the semiclassical approach\cite{Culcer2004a}.

\begin{figure*}
\includegraphics[width=6cm, height=4cm]{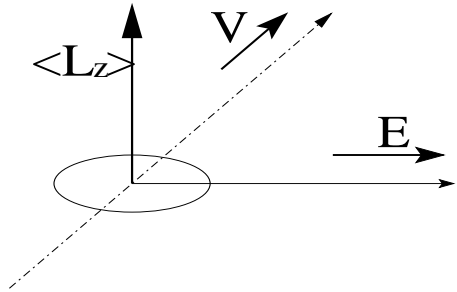}
\caption{ A simple sketch of  an orbital angular momentum flux.
The applied  electric field creates a force on the flux and force
it to drift in the perpendicular direction, which produces the
orbital current. } \label{fig1}
\end{figure*}

Several remarks are in order. First, as shown above, the
topological orbital angular momentum currents are the same in both
bands, unlike the spin Hall current which runs opposite in two
bands. Secondly, the orbital Hall current is a real topological
effect. If one compares it with the topological spin current in
the Luttinger case, the individual contribution to the orbital
Hall current from the single particle based on our effective
Hamiltonian is zero but the integral on the total Fermi surface is
non-vanishing, unlike the spin Hall current in which the
contribution from each single particle states is
non-vanishing\cite{Murakami2003}. A simple physical origin of the
orbital Hall current can be understood as follows. Since an
orbital angular momentum flux exists in the momentum space, by
applying a potential with an electric field, it feels the force
and drifts in the perpendicular direction to the force just like
the motion of the ordinary magnetic flux in real space. This
motion creates the current. A simple picture is sketched in
fig.\ref{fig1}. Thirdly, the topological orbital current should be
expected to be free from the disorder. The spin Hall current has
been proved to vanish even in the weak disorder limit. The reason
is simply that the effect of spin Hall current requires the
presence of both bands which is easily seen in the Kubo formula.
However, the topological nature of the orbital Hall current
maintains as long as the band gap at two Fermi surfaces is larger
than the energy scale of the disorder. Finally, when the Rashba
and Dresselhaus coupling strength are comparable, the orbital Hall
conductance can be very big. When the orbital Hall conductance is
too big, it is possible that a spontaneous magnetization takes
place inside the bulk.

In conclusion, a topological orbital angular momentum Hall current
exists in the Rashba-Dresselhaus spin orbit coupling systems. It
has an important part which is purely originated from the
existence of magnetic flux in momentum space in the adiabatic
limit. This orbital current is always larger than the spin Hall
current when both couplings are present. Therefore, the orbital
angular momentum current should be the major player in real
experiments. The prediction can be tested in the parameter region
when the Rashba coupling is larger than the Dresselhaus coupling
since the orbital and spin current runs in opposite directions.
The total accumulated magnetization at edges is expected to be
opposite to the prediction derived from a pure spin Hall current.

The author would like to thank L. Rokhinson for valuable
discussions and thank Y.S. Wu for correcting an error in our early
result. The author would also like to thank Chang Liu for reading
and discussing part of the work. This work is supported by Purdue
research funding.

%\bibliography{orbitalref}
%\bibliographystyle{h-physrev3}
%\nocite{*}

\end{document}